\documentclass[onecolumn,usenatbib,amsmath,amssymb,amsbsy]{mn2e}
\usepackage{amsmath}
\usepackage{amssymb}
\usepackage{natbib}
\usepackage{graphicx}
\def\gcc{\hbox{\rm\hskip.35em  g cm}$^{-3}$}
\def\rads{\hbox{\rm\hskip.35em  rad s}$^{-1}$}
\def\radss{\hbox{\rm\hskip.35em  rad s}$^{-2}$}
\def\cms{\hbox{\rm\hskip.35em  cm s}$^{-1}$}
\def\dcm{\hbox{\rm\hskip.35em  dyne cm}$^{-1}$}
\def\dcms{\hbox{\rm\hskip.35em  dyne cm}$^{2}$}
\def\ergcm{\hbox{\rm\hskip.35em  erg cm}$^{-1}$}
\def\gc{\hbox{\rm\hskip.35em  G cm}$^{2}$}

\newcommand{\apj}{ApJ}
\newcommand{\apjl}{ApJ}
\newcommand{\mnras}{MNRAS}
\newcommand{\nat}{Nature}

\newcommand{\apss}{Ap\&SS}

\newcommand{\prc}{Phys. Rev. C}
\newcommand{\prd}{Phys. Rev. D}

\begin{document}
\title[Microscopic Vortex Velocity for Neutron Stars]{Microscopic Vortex Velocity in the Inner Crust and Outer Core of Neutron Stars}

\author[G\"{u}gercino\u{g}lu \& Alpar]{Erbil G\"{u}gercino\u{g}lu$^{1}$\thanks{E-mail: egugercinoglu@gmail.com} and
 M. Ali Alpar$^{2}$\thanks{E-mail: alpar@sabanciuniv.edu}\\
$^{1}$Istanbul University, Faculty of Science, Department of Astronomy and Space Sciences, Beyaz{\i}t, 34119, Istanbul, Turkey \\
$^{2}$Sabanc{\i} University, Faculty of Engineering and Natural Sciences, Orhanl{\i}, 34956 Istanbul, Turkey \\
}

\date{Accepted . Received ; in original form}
\pagerange{\pageref{firstpage}--\pageref{lastpage}} \pubyear{2016}
\maketitle
\label{firstpage}

\begin{abstract}

Treatment of the vortex motion in the superfluids of the inner crust and the outer core of neutron stars is a key ingredient in modeling a number of pulsar phenomena, including glitches and magnetic field evolution. After recalculating the microscopic vortex velocity in the inner crust, we evaluate the velocity for the vortices in the outer core for the first time. The vortex motion between pinning sites is found to be substantially faster in the inner crust than in the outer core, $v_0^{\rm crust} \sim 10^{7}\mbox{\cms} \gg v_0^{\rm core} \sim 1\mbox{\cms}$. One immediate result is that vortex creep is always in the nonlinear regime in the outer core in contrast to the inner crust, where both nonlinear and linear regimes of vortex creep are possible. Other implications for pulsar glitches and magnetic field evolution are also presented.

\end{abstract}


\section{Introduction} 
\label{sec:introduction}

Like all dense, strongly interacting Fermi systems under a certain critical temperature, most parts of a neutron star are expected to be in superfluid states \citep{migdal59}. Observational evidence for neutron star superfluidity comes from the long recovery timescales following glitches \citep{baym69a} and more recently from the rapid cooling of the neutron star inside the CasA supernova remnant which indicates a transition into the superfluid/superconducting phase \citep{page11, shternin11}. A superfluid can achieve rotation only by forming quantized vortex lines. A neutron star's rotational dynamics is governed by the distribution and motion of these quantized vortex lines. Interaction of vortex lines with the ambient matter plays a significant role in the glitches \citep{alpar84a,ruderman98,sedrakian99}, thermal evolution \citep{alpar89, sedrakian93} and magnetic field evolution \citep{srinivasan90,jahan-miri00}. As the star spins down, the macroscopic rotation rate $\Omega_{\rm s}$ of the superfluid will follow the normal matter rotation rate $\Omega_{\rm c}$ at a lag $\omega=\Omega_{\rm s}-\Omega_{\rm c}$. In modeling the neutron star with a crust and a superfluid component, the equations of motion are
\begin{equation}
I_{\rm c}\dot\Omega_{\rm c}+I_{\rm s}\dot\Omega_{\rm s}=N_{\rm ext},
\end{equation}
and
\begin{equation}
\dot\Omega_{\rm s}=-\frac{2\Omega_{\rm s}v_{\rm r}(\omega)}{r},
\label{sspindown}
\end{equation}
where $N_{\rm ext}$ is the external braking torque, $I_{\rm c} (I_{\rm s})$ and $\dot\Omega_{\rm c} (\dot\Omega_{\rm s})$ are moment of inertia and spin-down rate of the crust (superfluid) component, respectively. The superfluid regions follow the spin-down of the neutron star's crust by sustaining a continuous vortex current in the radially outward direction with a rate 
\begin{equation}
v_{\rm r}= -\frac{r\dot\Omega}{2\Omega},
\label{steady}
\end{equation}
where $\dot\Omega$ and $\Omega$ are the spin-down and rotation rates of the pulsar respectively, and $r$ is the distance from the rotation axis. This steady state vortex motion corresponds to a steady state value of the lag $\omega$ attained as a result of the interactions of normal matter cores of the vortices with the components of the star that couple to the spin-down of the crust. Vortex cores interact with the electrons, the crustal lattice, the superconducting protons in the neutron star core and the quantized magnetic flux tubes of the proton superconductor.

We calculate the microscopic vortex velocity both in the inner crust and in the outer core by considering the Bernoulli force due to the excess kinetic energy of local induced superfluid flow around vortices arising from inhomogeneities presented by the nuclei and flux tubes. In Section \ref{sec:motion} we summarize the description of vortex motion in neutron stars. In Section \ref{sec:microscopicvel} we construct the basic formalism for obtaining the microscopic vortex velocity. In Section \ref{sec:crust} we reevaluate the microscopic vortex velocity in the inner crust while in Section \ref{sec:core} we obtain for the first time the microscopic vortex velocity in the outer core where the dynamics is determined by the interaction of the vortex lines with the quantized flux tubes of the proton superconductor. In Section \ref{sec:implication} we discuss the implications of our findings for pulsar glitches and magnetic field evolution. Section \ref{sec:conclusions} presents our conclusions.
\section{Description of Vortex Motion in Neutron Stars}
\label{sec:motion}

The equation of motion of a (straight) vortex moving
at velocity $\vec v_{\rm L}$  is determined from the balance between the Magnus response ``force"which depends on the relative velocity of the vortex with respect to the superfluid velocity $\vec v_{\rm s}$, and the physical forces acting on the vortex which depend on the vortex velocity with respect to the velocity $\vec v_{\rm c}$ of the normal matter corotating with the crust. These forces arise due to the interaction with lattice nuclei \citep{alpar77, epstein88} and phonons \citep{jones92} in the inner crust or with electrons \citep{alpar84b} and flux tubes \citep{sidery09} in the outer core. For drag forces linear in the velocity difference $\vec v_{\rm L}-\vec v_{\rm c}$, the equation of motion is
\begin{equation}
\rho_s \vec{\kappa} \times \left(\vec{v}_{\rm L}-\vec{v}_{\rm s} \right)-\eta (\vec{v}_{\rm c} - \vec{v}_{\rm L}) = 0, 
\label{drag}
\end{equation}
where $\rho_s$ is the superfluid mass density, $\kappa=h/2m_{\rm n}$ is the vorticity
quantum where $m_{\rm n}$ is the neutron mass and $h$ is Planck constant,  and $\eta$ is the drag coefficient. The $\vec{\kappa}$ vector is directed along
the vortex and parallel to the rotation axis. In cylindrical
coordinates $(r,\phi,z)$, with the rotation axis and $\vec{\kappa}$ in the $z$ direction, the vortex velocity is given by \citep{bildsten89}
\begin{equation}
\vec{v_L}=\omega R\left(\frac{1}{2}\sin
2\theta_d\,\hat{r}+\cos^2\theta_d\,\hat{\phi}\right),
\label{velocity}
\end{equation}
where $\omega= \Omega_{\rm s}-\Omega_{\rm c}$ is the angular velocity lag between the superfluid and the crust, and the dissipation angle $\theta_d$ is defined by 
\begin{equation}
\tan\theta_d\equiv\frac{\eta}{\rho_s\kappa}.
\end{equation}
Then a vortex moves at an angle $\theta_{\rm d}$ with respect to
the superfluid flow. Drag coefficients may differ by seven orders of magnitude for various processes \citep{haskell12, link14}. The drag coefficient and the dissipation angle are typically small, so that the vortex lines flow with a velocity close to the azimuthal macroscopic superfluid flow; $\vec v_{\rm L}\cong \vec v_{\rm s}=\omega r \hat \phi$, with a much smaller radial speed of the vortex lines, $v_{\rm r} \propto \eta$, whereby the drag force on the vortex lines spins down the superfluid (Eq. (\ref{steady})).

When the neutron superfluid is in a microscopically inhomogeneous medium, where the spacing between inhomogeneities is much less than the mean spacing between the vortex lines $l_{\rm v}=(2\Omega/ \kappa)^{-1/2}$, the forces determining the vortex motion are due to the local microscopic interaction with the inhomogeneities. In this situation, the magnitude of the vortex line velocity with respect to the normal matter, $v_0 \equiv \left|\vec v_{\rm L}-\vec v_{\rm c}\right|$, will not scale with the macroscopic average velocity difference $\omega R$ between the superfluid and the normal matter. The macroscopic average motion of the vortices, in particular their radial average speed $v_{\rm r}$ away from (towards) the rotation axis, that determines the spin-down (or spin-up) of the superfluid, is then related to a microscopic velocity $v_0$ in a statistical model. The directions of microscopic velocity are geometrically random, as determined by the distribution of inhomogeneities.

The vortex creep model \citep{alpar84a,alpar89} is a statistical model describing the macroscopic dynamics resulting from the vortex line interactions with the lattice of nuclei, at a lattice spacing $b \ll l_{\rm v}$, in the inner crust superfluid. The microscopic vortex velocity $v_0$ is employed to give a trial rate of vortex lines against potential pinning sites and barriers sustained by the nuclei. Thus, in the vortex creep model the vortex velocity is defined as a trial or microscopic random velocity $v_0$ times its rate in a preferred direction. Even though there is pinning, vortex lines can overcome pinning barriers due to the finite temperature $T$ and migrate radially outward as dictated by the external spin-down torque. This slow radial drift (``creep") rate is
\begin{equation} 
v_{\rm r}=2 v_0 e^{-E_{\rm p}/kT} \sinh\left(\frac{E_{\rm p}}{kT}\frac{\omega}{\omega_{\rm cr}} \right),
\label{creeprate}
\end{equation}
where $E_{\rm p}$ is the pinning energy and $\omega_{\rm cr}$ is the maximum angular velocity lag that can be maintained by pinning forces. The superfluid transfers angular momentum to the charged normal matter continuously in vortex creep, or in discrete glitch events via sporadic vortex discharges. The microscopic vortex velocity is a crucial parameter in determining the creep rate, and in particular whether creep has the full nonlinear dependence on the lag $\omega$. Comparing the steady state creep rate given by Eq.(\ref{steady}) with the model given in Eq.(\ref{creeprate}), one can decide whether the dependence on the lag $\omega$ is linear or nonlinear for a given pinning energy $E_{\rm p}$, temperature $T$ and microscopic vortex velocity $v_0$. Vortex creep will be in the linear (nonlinear) regime when $E_{\rm p}/kT$ is less (greater) than a transition value \citep{alpar89}:
\begin{equation}
\left( \frac{E_{\rm p}}{kT}\right)_{\rm tr}= \ln \left( \frac{4\Omega v_0}{|\dot\Omega|r}\right),
\label{transition}
\end{equation}
where $r \approx R_* \cong 10^{6}$cm is neutron star's radius. In the linear creep regime Eq.(\ref{sspindown}) becomes
\begin{equation}
\dot\Omega_{\rm s}=-\frac{\omega}{\tau_{\rm lin}}.
\end{equation}
Linear creep responds to perturbations by exponential relaxation with a timescale inversely proportional to $v_0$ \citep{alpar89}
\begin{equation}
\label{taulin}
\tau_{\rm lin} = \frac{k T }{E_{\rm p}} \frac{r \omega_{\rm cr}}{4 \Omega_{\rm s} v_{0}} \exp \left( \frac{E_{\rm p}}{kT} \right).
\end{equation}
In the nonlinear regime,
\begin{equation}
\sinh\left(\frac{E_{\rm p}}{kT}\frac{\omega}{\omega_{\rm cr}}\right)\cong \frac{1}{2}\exp\left(\frac{E_{\rm p}}{kT}\frac{\omega}{\omega_{\rm cr}}\right). \nonumber 
\end{equation}
The steady state lag in the nonlinear regime is 
\begin{equation}
\omega_{\infty}=\omega_{\rm cr}\left[1-\left(\frac{kT}{E_{\rm p}}\right) \ln\left(\frac{2\Omega_{\rm s}v{_0}}{|\dot\Omega|_{\infty}r}\right)\right]. \nonumber
\end{equation}
The post-glitch response of nonlinear creep is generally not simple exponential relaxation. Characteristic nonlinear response can be seen as the stopping of creep until a waiting time $t_{0}=\delta\omega/|\dot\Omega|_{\infty}$ determined from glitch induced change in the steady state lag and steady state spin down rate, or as a gradual power law recovery \citep{alpar84a,alpar89}  
\begin{equation}
\Delta\dot\Omega(t)=\Delta\dot\Omega(0)\left(1-\frac{t}{t_0}\right),
\label{nonlinear creep}
\end{equation}
where $\Delta\dot\Omega(0)$ is the glitch induced offset in the spin-down rate. As shown in Eq.(\ref{transition}), whether a given region of the superfluid is in the nonlinear or linear creep regime depends on the microscopic vortex velocity $v_0$ as well as on $E_{\rm p}, kT$ and other parameters. Although the dependence on $v_0$ is logarithmic, the range of the possibilities is wide. 
Both linear and nonlinear creep regimes exist in different parts of the neutron star superfluid. Pulsars exhibit very nonlinear post-glitch behaviour, as in Eq.(\ref{nonlinear creep}), along with simple exponential relaxation. 
\section{Determination of the Microscopic Vortex Velocity} 
\label{sec:microscopicvel}

The procedure for determining the microscopic vortex velocity $v_0$ is based on superfluid current conservation around a vortex while maintaining the quantized circulation $\kappa$. This is used to obtain the Bernoulli force which originates from the kinetic energy variation of local superluid flow around the vortex \citep{alpar77}. We will follow the highly simplified graphical description of \citet{alpar77} for the interaction of nuclei and vortices in the crust lattice. Later treatments employing the method of images \citep{shaham80} and employing the velocity field of a vortex against a nuclear potential in the complex plane \citep{epstein88} give similar results.
Taking into account the superfluid density difference inside and outside of an inhomogeneity (lattice nuclei in the inner crust, flux tubes in the outer core) current continuity and vorticity equations for the superfluid velocity around a vortex can be expressed as follows:
\begin{equation}
\label{vorcurrent}
\rho_{\rm in}v_{\rm in}(r)=\rho_{\rm out}v_{\rm out}(r),
\end{equation}
and
\begin{equation}
\label{vorticity}
r\left[\phi_{0}v_{\rm in}(r)+(2\pi-\phi_{0})v_{\rm out}(r)\right]=\kappa,
\end{equation}
where $\rho_{\rm in} (\rho_{\rm out})$ and $v_{\rm in} (v_{\rm out})$ are the superfluid density and the vortex velocity inside (outside) of the inhomogeneity, respectively and $\phi_0$ is the angle defining the angular size of the inhomogeneity as seen from the vortex axis. From Eqs. (\ref{vorcurrent}) and (\ref{vorticity}) we obtain:
\begin{equation}
v_{\rm in}(r)=\frac{\rho_{\rm out}}{\phi_{0}\rho_{\rm out}+(2\pi-\phi_{0})\rho_{\rm in}}\frac{\kappa}{r},
\label{vin}
\end{equation}

\begin{equation}
v_{\rm out}(r)=\frac{\rho_{\rm in}}{\phi_{0}\rho_{\rm out}+(2\pi-\phi_{0})\rho_{\rm in}}\frac{\kappa}{r}.
\label{vout}
\end{equation}
The Bernoulli force can be estimated by using a simple geometry. In our configuration adopted from \citet{alpar77}, an inhomogeneity with its center at a distance $R$ away from the vortex axis will affect the superfluid velocity field around a vortex only in a region bounded with $R-R_{\rm L}<r<R+R_{\rm L}$, $0<\phi<\frac{2R_{\rm L}}{R}$, $0<z<2R_{\rm L}$ in cylindrical coordinates centered on the vortex axis. Here $R_{\rm L}$ is the lengthscale, actually the effective radius, of the inhomogeneity. The kinetic energy increment due to this density inhomogeneity around a vortex is given by
\begin{eqnarray}
\Delta E &=& 2R_{\rm L} \int_{R-R_{\rm L}}^{R+R_{\rm L}}r dr \left( \int_{0}^{\phi_0} \frac{1}{2}\rho_{\rm in}v_{\rm in}^{2} (r) d \phi +\int_{\phi_0}^{2\pi} \frac{1}{2}\rho_{\rm out}v_{\rm out}^{2} (r) d \phi - \int_{0}^{2\pi} \frac{1}{2}\rho_{\rm out} \left[ \frac{\kappa}{2\pi r}\right]^{2} d \phi \right) \\ \nonumber
&=&R_{\rm L} \rho_{\rm out} \kappa^{2} \left ( \ln\frac{R+R_{\rm L}}{R-R_{\rm L}}\right) \left( \frac{\rho_{\rm in}R}{2R_{\rm L}(\rho_{\rm out}-\rho_{\rm in})+2\pi \rho_{\rm in}R} -\frac{1}{2\pi} \right).
\end{eqnarray}
The gradient of the above expression yields the Bernoulli Force:
\begin{align}
F_{\rm B }= -\frac{d \Delta E}{dr} 
= -R_{\rm L}\rho_{\rm out} \kappa^{2} \left[ \left( \ln \frac{R+R_{\rm L}}{R-R_{\rm L}}\right) \left( \frac{2\rho_{\rm in} \Delta \rho R_{\rm L}}{( 2R_{\rm L} \Delta \rho + 2\pi \rho_{\rm in} R)^{2}} \right) - \frac{2R_{\rm L}}{R^{2}-R_{\rm L}^{2}} \left(\frac{\rho_{\rm in} R}{2R_{\rm L} \Delta \rho + 2\pi \rho_{\rm in} R} - \frac{1}{2\pi} \right) \right],
\label{bernoulli}
\end{align}
where we defined $\Delta\rho=\rho_{\rm out}-\rho_{\rm in}$. The Bernoulli force appears because if the inhomogeneity is brought closer to the vortex the superfluid pressure will be lower on the side of the line where the induced circulation around it due to the density inhomogeneity adds constructively with the background flow, and higher on the opposite side. Thus, this force is attractive for $\rho_{\rm in} < \rho_{\rm out}$ and repulsive for $\rho_{\rm out} < \rho_{\rm in}$ \citep{alpar77, shaham80}. The resulting pressure gradient must be balanced with the Magnus response ``force" from which the microscopic vortex velocity is obtained as
\begin{equation}
\frac{F_{\rm B}}{R}=\rho \kappa v_0
\label{v0}
\end{equation}

Vortex lines will have typical velocities $v_0$ with respect to the background, average azimuthal flow of the superfluid. The direction of the vortex motion will depend on the dynamical position and orientation of the vortex line with respect to the inhomogeneities. Until recently, the setting for vortex creep was taken to be the inner crust of the neutron star, where nuclei in the crustal lattice can pin to the vortex lines of the neutron superfluid. This changed with the realization \citep{chamel05, chamel12} that the neutron effective mass in the crust lattice will be different from the bare mass, due to Bragg scattering of the neutrons. As a consequence of this ``entrainment effect" the crust superfluid may not provide enough mass and moment of inertia to explain the observed post-glitch relaxation \citep{chamel06, andersson12, chamel13}. In an earlier paper we pointed out that pinning and creep of vortex lines against toroidal flux tubes in the outer core can supply the needed additional component of the post-glitch relaxation \citep{erbil14}. The microscopic vortex velocity around flux tubes needs to be considered in this new context of pinning and creep against toroidal flux tubes in the outer core of the neutron star. In subsequent sections we will use Eq. (\ref{v0}) to deduce the $v_0$ value both in the inner crust and in the outer core, where the inhomogeneities posed by quantized flux tubes are treated.
\section{Vortex Velocity in the Inner Crust} 
\label{sec:crust}

The physical state of the inner crust with its neutron rich nuclei and dripped neutron superfluid interspersed with them is well established since the pioneering work of \citet{negele73}. The superfluid density inside a nucleus is found to be somewhat larger than the outside dripped superfluid, $\rho_{\rm in} > \rho_{\rm out}$, in most parts of the inner crust. This means that the superfluid velocity at points equidistant from the vortex axis is lowered inside a nucleus and becomes higher outside of it as compared to the homogeneous superfluid. This effect results in an increase in the total kinetic energy of the superfluid and thus brings about a Bernoulli force which keeps vortex lines away from the nucleus. For this case the relevant intersection lengthscale is the nuclear radius, $R_{\rm L}=R_{\rm N}$, the distance from vortex axis to the nucleus is the lattice constant, $R=b$, the opening angle is $\phi_0 = 2R_{\rm N}/b$ and in the densest pinning layer (where baryon density is $n_{\rm B}=7.89\times 10^{-2}$ fm$^{-3}$ and $\rho_{\rm in}=4.8 \times 10^{11}$ \gcc) with the aid of Eq.(\ref{bernoulli}) one obtains $\sim 1.2 \times 10^{18}$ \dcm for the Bernoulli force per unit length \citep{alpar77}. Eq.(\ref{v0}) gives the typical value of $v_0 = 10^{7}$ \cms for the microscopic vortex velocity in the inner crust. The Bernoulli force will be radial, and the vortex motion will be in tangential directions as dictated by the Magnus response ``force". In the inner crust regions where $\rho_{\rm out} > \rho_{\rm in}$, the Bernoulli force and vortex velocity directions will be reversed. In any case, the estimate of $v_0$ is based on a straight vortex line, interacting with a single nucleus. In reality the vortex will be bent and its motion is geometrically frustrated in the lattice, but $v_0$, as a trial rate at neighboring nuclei, is expected to lie within the estimated order of magnitude. Thus, vortex lines move comparatively fast between the pinning centers in the inner crust superfluid. Similar values of $v_0$ were estimated by \citet{shaham80} and \citet{epstein88}. Bragg scattering of dripped superfluid neutrons from lattice nuclei, the entrainment effect \citep{chamel05, chamel12}, does not have a significant effect on these calculations.

The value of $v_0$ is crucial in determining the workings of vortex creep, in particular whether the vortex creep is in the full nonlinear regime or in a linear regime. The rotational dynamics of the superfluid$-$normal matter system has a steady state in which the superfluid and normal matter spin down at the same rate at constant lag $\omega$. The steady state value of the macroscopic radial vortex flow rate $v_{\rm r}=\left|\dot\Omega\right|R/2\Omega$, together with the value of $v_0$, determines whether vortex creep is in the nonlinear regime, in which the response to a glitch induced offsets $\delta\omega$ in  $\omega$ is highly nonlinear, or in the linear regime, where the response is linear in $\delta\omega$, behaving as for drag forces with simple exponential relaxation. \citet{link14} assumed that once a vortex unpins it moves with the local angular velocity lag value $ \omega = \Omega_{\rm s} - \Omega_{\rm c} \lesssim 1$ \rads, taking the macroscopic azimuthal velocity $v_{\rm \phi}=\omega R\sim 10^{5}$\cms instead of the microscopic randomly oriented speed $v_0 \sim 10^{7}$\cms. This leads \citet{link14} to the conclusion that there is no linear creep regime within the entire inner crust. However, when the microscopic interactions are taken into account with $v_0 \sim 10^{7}$\cms one cannot rule out the possibility of linear creep regions in the inner crust.

Recently, \citet{haskell15} conducted numerical simulations of the vortex velocity in the inner crust by obtaining the lines' mean free path among adjacent pinning sites, i.e. crustal nuclei, with geometrical cross sections implicit. They found qualitative agreement with the vortex creep model of \citet{alpar84a}, which from the beginning takes into account of the microscopic random velocities $v_0 \sim 10^{7}$\cms and therefore includes both nonlinear and linear creep regimes.
\section{Vortex Velocity in the Outer Core}
\label{sec:core}

The microscopic velocity of vortices which is required to assess the creep motion of vortex lines against the flux tubes has not been evaluated properly before. Either the crustal value of $v_0 \sim 10^{7}$ \cms was used \citep{sidery09} or the expression that stems from the globally averaged value $v_{\rm \phi}=\omega R$ appropriate for homogeneous drag forces \citep{link14} was employed. Here we will make a rough estimate for the microscopic vortex velocity in the outer core by taking neutron star core physical circumstances into account. 

During the early stages of the neutron star's life, the proton phase transition from normal to superconducting fluid is accompanied by the formation of a mixed state for which magnetic flux is confined into discrete flux tubes with flux quantum $\Phi_{0}=hc/2e=2\times 10^{-7}$\gc  \citep{baym69b}. Numerical simulations in non-superfluid \citep{braithwaite09} and in superconducting \citep{lander12, lander14} canonical neutron stars show that for a stable magnetic field configuration inside neutron stars, a toroidal component of the magnetic field stronger than the surface field, localized in the outer layers of the core is necessary. For magnetars such a type II superconductivity is also expected \citep{lander14,fujisawa14}; the upper critical field for superconductivity is not exceeded since the Hall effect causes conversion of some part of the toroidal field's energy into the poloidal field and weakening of the interior toroidal field compared to the surface poloidal field. Due to the very high electrical conductivity of the neutron star core \citep{baym69c}, any stable magnetic field configuration will persist in equilibrium for a long time. However, flux tubes' interaction with the expanding vortex array in a spinning down neutron star may carry some magnetic flux out of the core. This possibly determines the long term magnetic and rotational evolution of the neutron star \citep{srinivasan90, jahan-miri00, jones06}.

In contrast to the poloidal field configuration, toroidal arrangement of the flux tubes offers topologically inevitable pinning sites for vortex lines and provides creep conditions similar to the inner crust \citep{sidery09, erbil14}. During their motion, the neutron superfluid's vortex lines will inevitably face intersections with toroidally oriented flux tubes. Two relevant lengthscales pertaining to the flux tubes, the magnetic field's London penetration depth $\Lambda_* $ and the distance $l_{\Phi}$ between flux tubes, are given by \citep{alpar84b,erbil14},
\begin{equation}
\Lambda_* \simeq 95 \left[\left( \frac{m_{\rm p}^{*}/m_{\rm p}}{0.5} \right) \left( \frac{x_{\rm p}}{0.05} \right)^{-1} \left(\frac{\rho}{10^{14}\mbox{\gcc} } \right)^{-1} \right]^{1/2} \rm fm,
\label{london}
\end{equation}
and
\begin{equation}
l_{\Phi}=\left(\frac{B_{\phi}}{\Phi_0}\right)^{-1/2}\simeq 450\left(\frac{B_{\phi}}{10^{14} \rm G}\right)^{-1/2} \rm fm,
\label{invortex}
\end{equation}
where $x_{\rm p}$ is the proton fraction, $m_{\rm p}$  and $m_{\rm p}^*$ are proton bare and effective mass respectively, and $B_{\phi}$ is the toroidal component of the magnetic field.

For $r \leq \Lambda_*$ the ambient pressure of the neutron star matter is partially screened by the magnetic and Bernoulli pressures associated with the flux tube, leading to a pressure drop in the region $\xi_{\rm p} \leq r \leq \Lambda_*$ ($\xi_{\rm p}$, the coherence length, being flux tube core radius) from the flux tube axis. The pressure drop inside a flux tube is \citep{muslimov85,wendell88}
\begin{equation}
\Delta P(r)= \frac{H^{2}(r)}{8\pi}+\frac{1}{2}\rho_{\rm p}v_{\rm p}^{2}(r) \simeq \frac{1}{8\pi} \left[ \frac{\Phi_0}{2\pi \Lambda_{*}^{2}}\ln \left( \frac{\Lambda_*}{r}\right) \right]^{2}+\frac{1}{2}\rho_{\rm p}v_{\rm p}^{2}(r),
\label{pressure}
\end{equation}
and this pressure drop causes a small decrement in the surrounding density
\begin{equation}
\Delta \rho (r) \simeq \frac{d \rho}{d P}=\frac{\rho}{\Gamma P} \Delta P(r),
\label{density}
\end{equation}
where $\Gamma$ is the adiabatic index and $v_{\rm p}=\kappa /2 \pi r$ is the velocity field around a flux tube. We will calculate the density difference at $r= \xi_{\rm p}$. The coherence length is \citep{mendell91}
\begin{equation}
\xi_{\rm p}=16x_{\rm p}^{1/3}\rho_{14}^{1/3} \frac{m_{\rm p}}{m_{\rm p}^{*}} \Delta_{\rm p} (\rm MeV)^{-1} \rm fm,
\label{coherence}
\end{equation}
where $\rho_{14}$ is the density in terms of $10^{14}$\gcc, $\Delta_{\rm p}$ is the proton pairing energy gap. To find $\Delta \rho / \rho$ we exploit the equation of state parameters from \citet{akmal98} and proton superconductor parameters from \citet{baldo07}: $\rho \approx 2 \times 10^{14}$\gcc, $P \approx 2.035 \times 10^{33}$\dcms, $x_{\rm p}=0.041$, $\Gamma \approx 2.7$, $\Delta_{\rm p} \approx 1.2$ MeV, $m_{\rm p}^* /m_{\rm p} \approx 0.9$. With these numerical values in Eqs. (\ref{pressure})-(\ref{coherence}) we arrive at $\Delta \rho / \rho \cong 1.5 \times 10^{-5}$. For the outer core conditions the relevant flux tube$-$vortex line intersection lengthscale is the London penetration depth, $R_{\rm L}=\Lambda_*$, the average distance from vortex axis to the flux tube is of the order of the flux tube separation, $R=l_{\Phi}$, and the opening angle is $\phi_0 = 2 \Lambda_{*}/l_{\Phi}$. In the end by using Eqs. (\ref{bernoulli})-(\ref{invortex}) we find $v_{0} \simeq 0.7$\cms for microscopic vortex velocity in the outer core. The reasons for such a low microscopic velocity in the outer core compared to the corresponding value in the inner crust are;
\begin{enumerate}
\item the difference in the superfluid density contrast between inside and outside of inhomogeneities (nuclei or flux tubes) and especially,
\item the large difference between the respective  lengthscales, the lattice spacing in the inner crust $b$, and the spacing of toroidal flux tubes  $l_{\Phi}$ in the outer core.
\end{enumerate}

From the value of $v_0$ one can discriminate whether vortex creep against toroidal flux tubes is in the linear or nonlinear regime. The linear to nonlinear creep transition is determined from Eq.(\ref{transition}). Typical parameters for the Vela pulsar, interior temperature $T \approx 10^{8}$K, spin-down rate $\vert \dot\Omega \vert \cong 10^{-10}$\radss, angular velocity $\Omega \cong 70$\rads, give $E_{\rm p}|_{\rm tr}=0.12$ MeV for linear to nonlinear creep transition value of pinning energy. There are two estimates for the pinning energy of vortex$-$flux tube junctions in the literature, one accounts for change in the condensation energies of vortex and flux tube cores and the other considers the interaction energy contained in magnetic fields of the concurrent structures. 
The pinning energy arising from proton density fluctuations is \citep{muslimov85,sauls89}
\begin{equation}
E_{\rm p} = \frac{3}{8}n_{\rm n} \frac{\Delta_{\rm p}^{2}}{E_{\rm F_{\rm p}}^{2}} \frac{\Delta_{\rm n}^{2}}{E_{\rm F_{\rm n}}} \left( \xi_{\rm n}^{2} \xi_{\rm p} \right) = \frac{1}{\pi^{5}}\frac{\Delta_{\rm p}}{x_{\rm p}} \left( \frac{m_{\rm n}^{*}}{m_{\rm n}}\right)^{-2} \left( \frac{m_{\rm p}^{*}}{m_{\rm p}}\right)^{-1} \simeq 0.13 {\rm MeV} \left( \frac{\Delta_{\rm p}}{1 {\rm MeV}}\right) \left( \frac{x_{\rm p}}{0.05}\right)^{-1} \left( \frac{m_{\rm n}^{*}/m_{\rm n}}{1}\right)^{-2} \left( \frac{m_{\rm p}^{*}/m_{\rm p}}{0.5}\right)^{-1},
\label{pincond}
\end{equation}
where $\Delta_{\rm p}$ is the proton pairing energy gap, $E_{\rm F}$ is Fermi energy, $\xi$ is coherence length, $m^*/m$ is the effective to bare mass ratio, subscripts ``n" and ``p" refer to neutrons and protons, respectively and $n_{\rm n}$ is the neutron number density.
In the neutron star core, vortex lines are strongly magnetized due to proton supercurrents dragging around them. This endows each vortex line with a field intensity $B_{\rm v}$ comparable to that of a flux tube $B_{\Phi} \sim 10^{15}$G \citep{sedrakian83,alpar84b}. The pinning energy contribution coming from the overlap of the magnetic fields of the flux tube$-$vortex line reads \citep{mendell91, jones91, chau92} \footnote{Note that there is an unfortunate typo in \citet{ruderman98} (propagated in \citet{link12}) of a factor $\pi$ rather than $1/\pi$ in the magnetic energy. This leads to a factor of $\pi^{2}\sim10$ times larger pinning energies.}
\begin{equation}
E_{\rm p} = \frac{2\vec{B}_{\rm V}\cdot \vec{B}_{\Phi}}{8\pi}\left(\pi \Lambda_*^{2} \ell_{\Lambda} \right) = \frac{\Phi_0^{2}}{16\pi^{2}}\frac{\ell_{\Lambda}}{\Lambda_*^{2}}\frac{\delta m_{\rm p}^{*}}{m_{\rm p}^{*}} \ln\left(\frac{\Lambda_*}{\xi_{\rm p}}\right) \cos \theta,
\label{pinmag}
\end{equation}
where $\ell_{\Lambda}$ and $\theta$ denote the overlap length and the angle between the flux tube and the vortex line, respectively, $\delta m_{\rm p}^{*}  =\vert m_{\rm p}^{*}-m_{\rm p}\vert$. 

For a simple geometry, the overlap length can be expressed in terms of the London penetration depth and the angle between a flux tube and a vortex line as follows:
\begin{equation}
\ell_{\Lambda}\simeq \frac{2 \Lambda_*}{\sin \theta}.
\label{overlap}
\end{equation}
As neither the flux tube nor the vortex line have infinite rigidity, both structures can bend at the junction. A vortex line has a finite energy per unit length (tension) given by
\begin{equation}
T_{\rm v}= \frac{\rho_{\rm s}\kappa^{2}}{4\pi}\ln \frac{l_{\rm v}}{\xi_{\rm n}}.
\end{equation}
This originates from the kinetic energy associated with the velocity field $\propto r^{-1}$ around the vortex line. In terms of typical parameters vortex line tension in neutron stars is \citep{andersson07}
\begin{equation}
T_{\rm v} \simeq 10 ^{9} \left(\frac{\rho_{\rm s}}{2 \times 10^{14}\mbox{\gcc}}\right) \mbox{\ergcm}, 
\label{vorten}
\end{equation}
with
\begin{equation}
\ln \frac{l_{\rm v}}{\xi_{\rm n}} \approx 20-\frac{1}{2}\ln \left(\frac{\Omega}{100 \mbox{\rads}}\right).
\end{equation} 
On the other hand, the flux tube tension is given by \citep{harvey86}
\begin{equation}
T_{\Phi}=\left(\frac{\Phi_0}{4\pi \Lambda_*}\right)^{2}\ln\left(\frac{\Lambda_*}{\xi_{\rm p}}\right) \sim 10^{7}\left(\frac{m_{\rm p}^{*}/m_{\rm p}}{0.5}\right)^{-1}\left(\frac{x_{\rm p}}{0.05}\right)\left(\frac{\rho_{\rm s}}{2\times 10^{14} \mbox{\gcc}}\right) \mbox{\ergcm}.
\label{fluxten}
\end{equation}
As can be seen from Eqs. (\ref{vorten}) and (\ref{fluxten}) a vortex line is $\sim 100$ times stiffer than a flux tube. When a vortex line and a flux tube come closer and intersect we can safely assume that the vortex line remains almost straight while the flux tube bends and twists. Some part of the energy gained by the overlapping of a vortex line with a flux tube goes over to the flux tube's lengthening during this bending process. When this effect is taken into account, the net energy gain becomes
\begin{equation}
E_{+}=E_{\rm p}-\Delta \ell_{\Phi}T_{\Phi}.
\label{egain}
\end{equation}
The lengthening of the flux tube around a vortex line amounts to
\begin{equation}
\ell_{\Phi} \approx 2\Lambda_* \left(\frac{1}{\sin \theta}-1\right),
\label{lengthen}
\end{equation}
since the attraction range is the circulating supercurrent's lengthscale $\Lambda_*$. 
To find the value of $\theta$ that makes energy gain maximum, we substitute Eqs. (\ref{pinmag}), (\ref{overlap}), (\ref{fluxten}), (\ref{lengthen}) in Eq.(\ref{egain}) for $E_{+}$ and then vary the ensuing expression with respect to $\theta$ to get
\begin{equation}
\theta=\arccos\left(\frac{1}{2}\frac{\delta m_{\rm p}^{*}}{m_{\rm p}}\right).
\end{equation}

Thus, the net energy gain arising from the vortex line$-$flux tube magnetic pinning becomes
\begin{eqnarray}
E_{+}&=\frac{\Phi_0^2}{8\pi^2 \Lambda_*}\ln\left(\frac{\Lambda_*}{\xi_{\rm p}}\right)\left[\left(\frac{\delta m_{\rm p}^{*}}{ m_{\rm p}}\right)\cot\left(\arccos\left(\frac{1}{2}\frac{\delta m_{\rm p}^{*}}{m_{\rm p}}\right)\right)-\frac{2}{\sin\left(\arccos\left(\frac{1}{2}\frac{\delta m_{\rm p}^{*}}{m_{\rm p}}\right)\right)}+2\right] \approx  4.8~  \mbox{MeV}\left(\frac{m_{\rm p}^{*}/m_{\rm p}}{0.5}\right)^{-1/2}\left(\frac{x_{\rm p}}{0.05}\right)^{1/2}\times \nonumber \\
&\times\left(\frac{\rho_{\rm s}}{2\times 10^{14} \mbox{\gcc}}\right)^{1/2}\left[\left(\frac{\delta m_{\rm p}^{*}/m_{\rm p}}{0.5 }\right)\cot\left(\arccos\left(\frac{1}{2}\frac{\delta m_{\rm p}^{*}/m_{\rm p}}{0.5}\right)\right)-\frac{2}{\sin\left(\arccos\left(\frac{1}{2}\frac{\delta m_{\rm p}^{*}/m_{\rm p}}{0.5}\right)\right)}+2\right].
\label{netgain}
\end{eqnarray}
This can be compared with the magnetic pinning energy estimate used earlier in the literature, i.e. Eq.(\ref{pinmag}), which does not take properly flux tube bending and lengthening into account. With   
$\ell_{\Lambda}=2\Lambda_*$ and $\cos\theta=1$ one obtains for Eq.(\ref{pinmag}),
\begin{equation}
E_{\rm p} \approx 40 \mbox{MeV}\left(\frac{\delta m_{\rm p}^{*}/m_{\rm p}}{0.5}\right)\left(\frac{m_{\rm p}^{*}/m_{\rm p}}{0.5}\right)^{-1/2}\left(\frac{x_{\rm p}}{0.05}\right)^{1/2}\left(\frac{\rho_{\rm s}}{2\times 10^{14} \mbox{\gcc}}\right)^{1/2}.
\end{equation}

Thus, the energy disposed for flux tube bending reduces the magnetic pinning energy estimate by a factor $\sim 8$. In order to determine net pinning energy gain in toroidal field region close to the crust$-$core interface, we use once again the equation of state parameters from \citet{akmal98} and proton superconductor parameters from \citet{baldo07}: $\rho \approx 2 \times 10^{14}$\gcc, $x_{\rm p}=0.041$, $\Delta_{\rm p} \approx 1.2$ MeV, $m_{\rm p}^* /m_{\rm p} \approx 0.9$. These numerical values give  $E_{+}\simeq 0.5$ MeV. For both of the pinning energy estimates given in Eq. (\ref{pincond}) and for the realistic case with magnetic and bending energies are taken into account in (\ref{netgain}), $E_{\rm p}> E_{\rm p}|_{\rm tr}=0.12$ MeV. We conclude that the vortex creep across the toroidal flux tube configuration is always in the nonlinear regime, as expected from the postglitch relaxation analysis of \citet{erbil14}. 

The critical angular velocity lag that can be sustained by pinning forces is found by equating  the pinning force to the Magnus ``response" force, $E_{+}/(\Lambda_{*}l_{\Phi})=\rho_{\rm s}\kappa R_{*} \omega_{\rm cr}$, and is given by
\begin{equation}
\omega_{\rm cr}\cong 6.3\times 10^{-2} \mbox{\rads}\left(\frac{\delta m_{\rm p}^{*}/m_{\rm p}}{0.5}\right)\left(\frac{m_{\rm p}^{*}/m_{\rm p}}{0.5}\right)^{-1}\left(\frac{x_{\rm p}}{0.05}\right)\left(\frac{B_{\phi}}{ 10^{14} \mbox{G}}\right)^{1/2}.
\end{equation}
In the non-linear creep regime vortex lines migrate radially outwards with a steady state angular velocity lag $\omega_{\infty}$, which is related to the the critical lag $\omega_{\rm cr}$   
\begin{equation}
\omega_{\rm cr}-\omega_{\infty}=\frac{kT}{\rho \kappa R_{*}\Lambda_{*} l_{\Phi}}\ln \left(\frac{2\Omega_{\rm s}v_{0}}{\left|\dot\Omega\right|R_{*}}\right).
\label{unpinning}
\end{equation}
\section{Implications for Pulsar Glitches and Magnetic Field Evolution}
\label{sec:implication}

Application of the creep model to the Vela and the Crab glitches \citep{alpar84a, alpar96} suggests that crustquakes may be triggering the glitches. The superfluid plays the role of amplifying the glitch event to the observed magnitude via sudden unpinning of a large number of the vortex lines in the inner crust. Since these vortices travel radially outward, the vortex lines within the toroidal field region in the outer core will not participate in making the glitch but will contribute to the post-glitch relaxation in response to the offset in the lag due to the change in the rotational state of the crust $\Delta\Omega_{\rm c}$. Creep in the toroidal field region not only meets the requirement of extra moment of inertia to resolve the problem raised by entrainment but also fits the post-glitch behavior of the Vela pulsar \citep{erbil14} and of pulsars of different ages \citep{erbil16a}.

Exceptionally large glitches in PSR B2334+61 \citep{yuan10} and PSR J1718--3718 \citep{manchester11} with magnitudes $\Delta\Omega_{\rm c}/\Omega_{\rm c} \gtrsim 2\times10^{-5}$ requires involvement of the toroidal field region in the glitch event itself. Evaluating Eq.(\ref{unpinning}) for vortex creep in the toroidal field region, we find that, here the steady state lag $\omega_{\infty}$ is much closer to the critical lag $\omega_{\rm cr}$, which is the threshold for an unpinning avalanche, than is the case in the crust superfluid:
\begin{equation}
\frac{\left(\omega_{\rm cr}-\omega_{\infty}\right)_{\rm crust}}{\left(\omega_{\rm cr}-\omega_{\infty}\right)_{\rm core}}\approx\frac{\Lambda_{*} l_{\Phi}}{b\xi_{\rm n}}\frac{\ln \left(\frac{2\Omega_{\rm s}v_{0}^{\rm crust}}{\left|\dot\Omega\right|R_{*}}\right)}{\ln \left(\frac{2\Omega_{\rm s}v_{0}^{\rm core}}{\left|\dot\Omega\right|R_{*}}\right)}\sim 200 \left(\frac{\Lambda_{*}}{100 \mbox{ fm}}\right)\left(\frac{l_{\Phi}}{500 \mbox{ fm}}\right)\left(\frac{b}{50 \mbox{ fm}}\right)^{-1}\left(\frac{\xi_{\rm n}}{10 \mbox{ fm}}\right)^{-1}.
\label{ratiounpin}
\end{equation}
Here the neutron star interior is taken to be isothermal from the inner crust superfluid into the core. We take $\Omega_{\rm s} = 100 \mbox{\rads},|\dot\Omega|=10^{-10} \mbox{\radss}, r=R_{*}\cong10^{6}$ cm values in the logarithmic expressions. It is seen that the difference between steady state lag and the critical lag for unpinning in the outer core is two orders of magnitude less than the case in the inner crust. In typical Vela and other pulsar glitches of magnitude $\Delta\Omega_{\rm c}/\Omega_{\rm c}\sim 10^{-6}$, the unpinning event starts in the crust superfluid, with the help of a crustquake acting as trigger. The exceptionally large glitches may have started in the toroidal flux region, as self organized events involving the vortex unpinning and creep process only. Possible superfluid based triggers include r-modes \citep{glampedakis09} and superfluid turbulence \citep{peralta06, erbil16b} for large scale self organized vortex unpinning. If these mechanisms, effective in the core superfluid, are involved, the largest glitches should start from these deepest pinned superfluid regions as predicted by \citet{sidery09}. Despite the relative closeness of $\omega_{\infty}$ and $\omega_{\rm cr}$, such events proceeding from the core are still rare. The reasons for very large glitches like in PSR B2334+61 \citep{yuan10} and PSR J1718--3718 \citep{manchester11} being rare may be particularly favorable relative orientation or radial extension of the toroidal field region with respect to the rotational axis in these pulsars.

This discussion of vortex microscopic motion and creep against flux tubes is based on the expectation that the flux tubes are stationary in the crust and normal matter frame of reference, to a good approximation.
As a result of the low microscopic vortex velocity in the outer core, presumably vortex lines cannot push the flux tube network enough and this may delay the magnetic flux expulsion from the core. Let us elaborate on this point. The vortex lines move outwards through the core in response to the spin-down of the normal matter, mediated by various forces on them. The resulting effective force that the vortices exert on flux tubes was dubbed the "vortex acting force`` by \citet{ding93}. This force tends to accelerate the magnetic field decay via motion of flux tubes with a magnitude in proportion to the ratio of number densities of vortex lines to the flux tubes \citep{ding93,jahan-miri00}
\begin{equation}
\vec F_{\rm n}=\frac{n_{\rm v}}{n_{\Phi}}\vec F_{\rm M}=\frac{2\Phi_{0}\rho R_{\rm core} \Omega_{\rm s}(t)\omega(t)}{B_{\rm core}(t)}\hat e_{\rm r},
\end{equation}
where $R_{\rm core}\approx R_{*}$ is radius of the location of the crust--core interface. In this expression it is implicitly assumed that forces on vortex lines are instantaneously communicated to the flux tube array. These authors estimated the force per unit length on a flux tube due to its interactions with vortex lines as the force that sustains the macroscopic velocity lag $\omega_{\infty}R$ between the superfluid and the normal matter via the Magnus effect. In reality, the relative velocity between flux tubes and vortex lines is the microscopic velocity $v_{0}^{\rm core}\cong 1\mbox{\cms}$ sustained by the local Bernoulli force, which is much less than the macroscopic velocity difference $\omega_{\infty}R\cong 10^{4}\mbox{\cms}$. This leads to a much lower estimate of the force per unit length of the flux tube applied by vortex lines by a factor of $(v_{0}^{\rm core}/\omega_{\infty}R)\sim10^{-4}$. As a result, the effect of vortex lines is negligible on the relaxation of the global magnetic field configuration: 
\begin{equation}
F_{\rm v-\Phi}\approx \frac{2\Phi_{0}\rho \Omega_{\rm s}(t)v_{0}^{\rm core}}{B_{\rm core}(t)}\sim8 \times 10^{-3}\left(\frac{\Omega_{\rm s}}{100 \mbox{\rads}}\right)\left(\frac{\rho_{\rm s}}{2\times10^{14}\mbox{\gcc}}\right)\left(\frac{B_{\rm core}}{ 10^{12} \mbox{ G}}\right)^{-1}\left(\frac{v_{0}^{\rm core}}{ 1 \mbox{\cms}}\right)\mbox{\dcm}
\end{equation}
Thus, with our estimate of the microscopic velocity $v_{0}^{\rm core}$ in the toroidal field line region, the effect of vortex lines in the dynamics of the flux tubes is negligible in comparison to magneto-hydrodynamic forces which typically have magnitudes $F\lesssim 10\mbox{\dcm}$ \citep{jahan-miri00}. This result is important, as it justifies the implicit assumption of \citet{ruderman98}, who compares the secular relaxation times of the vortices and the flux tubes on the basis of separate dynamics, neglecting the interaction between the two systems. During vortex creep the flux tubes can be taken to be stationary in the frame of charged matter and normal crust.

The poloidal and toroidal flux tube configuration, like any magnetic field configuration, has a tendency to relax by diffusion, buoyancy and other effects. \citet{jones06} calculated the flux tube velocity in connection with the bulk force resulting from the divergence of the stress tensor for the type II proton superconductor and obtained $v_{\Phi}\approx 4\times 10^{-7} \mbox{\cms}$. He assumed only poloidal configuration for flux tubes and neglected interactions with the surrounding vortices. Such a high flux tube velocity would result in a complete magnetic field expulsion from the entire core in a rather short time scale $\tau_{\rm decay}\approx R_{\rm core}/v_{\Phi}\sim 10^{5}\mbox{ yrs}\lesssim \tau_{\rm ohmic}$ comparable to the Ohmic dissipation timescale of the crustal currents. However, the stabilizing effect of the toroidal field arrangement of the flux tubes as well as the vortex lines kept within the same region through pinning and creep will resist further field decay. Thus, in the end a residual magnetic field that does not diffuse out will be expected at the late stages of the evolution. Note also that in the creep process the time at which a single vortex line remained pinned to numerous flux tubes is of the order of $2\Lambda_*/v_0 \sim 10^{-11}$s at each encounter so that the amount of the magnetic flux tubes carried by vortices during spin-down might be smaller than that predicted by \citet{srinivasan90}. The implications of these results on the flux tube--vortex line interaction, regarding flux expulsion induced by spin-down on evolutionary timescales will be considered in a separate work.
\section{Conclusions} \label{sec:conclusions}

In the vortex creep model the mean creep rate is simply a microscopic vortex velocity $v_0$ times the jump rate in a preferred direction which is radially outward as dictated by the external spin-down torque. For a vortex line in an inhomogeneous medium, the microscopic vortex velocity stems from Bernoulli forces caused by inhomogeneities within the superfluid, with the conditions that superfluid current circulation around a vortex line must be continuous and total vorticity should be equal to vorticity $\kappa$. For the inner crust superfluid, the  presence of nuclei leads to different vortex velocity fields inside and outside of the region bounded by the coherence length. As a consequence, the superfluid's kinetic energy differs from that of a homogeneous superfluid. With the extra kinetic energy vortices experience a Bernoulli force and gain a microscopic vortex velocity. \citet{alpar77} found  $v_0 \sim 10^{7}$  \cms for the inner crust circumstances which we reproduced here. \citet{sidery09} assumed the inner crust value  $v_0 \sim 10^{7}$  \cms for the neutron star core, despite the different physical circumstances. For the outer core superfluid, the presence of flux tubes leads to different vortex velocity fields inside and outside of the region limited by the London penetration depth. We have presented the first calculation of the microscopic velocity of the vortex lines which are creeping against the toroidal arrangement of flux tubes in the outer core. This velocity is found to be significantly smaller than the crustal value,  $v_0^{\rm crust} \sim 10^{7}\mbox{\cms} \gg v_0^{\rm core} \sim 1\mbox{\cms}$. Due to the very low microscopic velocity of vortex lines interspersed with flux tubes, the vortex creep against toroidal flux tubes will always be in the nonlinear regime as \citet{erbil14} have predicted. For vortex line--flux tube pinning we take flux tube bending into account and obtain an order of magnitude reduction for the resulting pinning energy compared to the estimates used earlier in the literature. Our findings have bearing conclusions for pulsar glitches and magnetic field decay from the neutron star core. The size of the pinning center and the range of the pinning interaction are both larger in the outer core than they are in the inner crust, $\Lambda_{*}> \xi_{\rm n}$ and $l_{\Phi}> b$, respectively. As a consequence, local fluctuations can raise flow of vortex lines relative to the background steady state creep rate $\omega_{\infty}$ to $\omega_{\rm cr}$ above which vortex discharges occur rather easily in the neutron star core compared to the inner crust. We speculate that fluid instabilities related to r modes or superfluid turbulence may initiate the largest glitches in the toroidal field region which is the innermost pinning region inside the neutron star. Due to the very low vortex velocity vicinity of a flux tube, vortex lines cannot push flux tubes array enough. Also since there is no perfect pinning inside neutron stars, vortex lines cannot carry most flux tubes with them during their motion. In a future paper we plan to address the problem of flux tubes' expulsion coupled to the motion of vortex lines and spin-down on evolutionary timescales.

\section*{acknowledgements}

We are thankful to the referee for useful criticism. This work is supported by the Scientific and Technological Research Council of Turkey (T\"{U}B\.{I}TAK) under the grant 113F354. M.~A. Alpar
is a member of the Science Academy (Bilim Akademisi), Turkey. 


\label{lastpage}

\end{document}